\pdfoutput=1
\documentclass{beatcs}

\usepackage{newunicodechar}
\usepackage{silence}
\newunicodechar{♢}{\tikz \node[inner sep=1.5,draw,diamond] {};}
\newunicodechar{☆}{\tikz \node[inner sep=1,draw,star,star point ratio=2] {};}
\newunicodechar{△}{\triangle}
\newunicodechar{⬜}{\kern 0.5pt\tikz \node[inner sep=1.7,draw,regular polygon,regular polygon sides=4] {};\kern 0.5pt}
\newunicodechar{◯}{\tikz[baseline=-3pt] \node[inner sep=1.7,draw,cloud,cloud puffs=4,cloud puff arc=190] {};}

\newunicodechar{⊥}{\bot}
\newunicodechar{•}{\item}
\newunicodechar{✓}{\checkmark}
\newunicodechar{✗}{\xmark}
\newunicodechar{…}{\dots}
\newunicodechar{≔}{\coloneqq}
\newunicodechar{⁻}{^-}
\newunicodechar{⁺}{^+}
\newunicodechar{₋}{_-}
\newunicodechar{₊}{_+}
\newunicodechar{ℓ}{\ell}
\newunicodechar{•}{\item}
\newunicodechar{…}{\dots}
\newunicodechar{≔}{\coloneqq}
\newif\ifnormopen\normopenfalse
\newunicodechar{‖}{\ifnormopen\rVert\normopenfalse\else\rVert\normopentrue\fi}
\newunicodechar{≤}{\leq}
\newunicodechar{≥}{\geq}
\newunicodechar{≰}{\nleq}
\newunicodechar{≱}{\ngeq}
\newunicodechar{⊕}{\oplus}
\newunicodechar{⊗}{\otimes}
\newunicodechar{≠}{\neq}
\newunicodechar{¬}{\neg}
\newunicodechar{≡}{\equiv}
\newunicodechar{₀}{_0}
\newunicodechar{₁}{_1}
\newunicodechar{₂}{_2}
\newunicodechar{₃}{_3}
\newunicodechar{₄}{_4}
\newunicodechar{₅}{_5}
\newunicodechar{₆}{_6}
\newunicodechar{₇}{_7}
\newunicodechar{₈}{_8}
\newunicodechar{₉}{_9}
\newunicodechar{⁰}{^0}
\newunicodechar{¹}{^1}
\newunicodechar{²}{^2}
\newunicodechar{³}{^3}
\newunicodechar{⁴}{^4}
\newunicodechar{⁵}{^5}
\newunicodechar{⁶}{^6}
\newunicodechar{⁷}{^7}
\newunicodechar{⁸}{^8}
\newunicodechar{⁹}{^9}
\newunicodechar{∈}{\in}
\newunicodechar{∉}{\notin}
\newunicodechar{⊂}{\subset}
\newunicodechar{⊃}{\supset}
\newunicodechar{⊆}{\subseteq}
\newunicodechar{⊇}{\supseteq}
\newunicodechar{⊄}{\nsubset}
\newunicodechar{⊅}{\nsupset}
\newunicodechar{⊈}{\nsubseteq}
\newunicodechar{⊉}{\nsupseteq}
\newunicodechar{∪}{\cup}
\newunicodechar{∩}{\cap}
\newunicodechar{∀}{\forall}
\newunicodechar{∃}{\exists}
\newunicodechar{∄}{\nexists}
\newunicodechar{∨}{\vee}
\newunicodechar{∧}{\wedge}
\newunicodechar{ℝ}{\mathbb{R}}
\newunicodechar{ℕ}{\mathbb{N}}
\newunicodechar{𝔼}{\mathbb{E}}
\newunicodechar{𝔽}{\mathbb{F}}
\newunicodechar{ℤ}{\mathbb{Z}}
\newunicodechar{𝒪}{\mathcal{O}}
\newunicodechar{⌊}{\lfloor}
\newunicodechar{⌋}{\rfloor}
\newunicodechar{⌈}{\lceil}
\newunicodechar{⌉}{\rceil}
\newunicodechar{·}{\cdot}
\newunicodechar{∘}{\circ}
\newunicodechar{×}{\times}
\newunicodechar{↑}{\uparrow}
\newunicodechar{↓}{\downarrow}
\newunicodechar{→}{\rightarrow}
\newunicodechar{←}{\leftarrow}
\newunicodechar{⇒}{\Rightarrow}
\newunicodechar{⇐}{\Leftarrow}
\newunicodechar{↔}{\leftrightarrow}
\newunicodechar{⇔}{\Leftrightarrow}
\newunicodechar{↦}{\mapsto}
\newunicodechar{∅}{\varnothing}
\newunicodechar{∞}{\infty}
\newunicodechar{≅}{\cong}
\newunicodechar{≈}{\approx}
\newunicodechar{ℓ}{\ell}
\newunicodechar{𝟙}{\mathds{1}}
\newunicodechar{𝟘}{\mathds{0}}

\newunicodechar{α}{\alpha}
\newunicodechar{β}{\beta}
\newunicodechar{γ}{\gamma}
\newunicodechar{Γ}{\Gamma}
\newunicodechar{δ}{\delta}
\newunicodechar{Δ}{\Delta}
\newunicodechar{ε}{\varepsilon}
\newunicodechar{ζ}{\zeta}
\newunicodechar{η}{\eta}
\newunicodechar{θ}{\theta}
\newunicodechar{Θ}{\Theta}
\newunicodechar{ι}{\iota}
\newunicodechar{κ}{\kappa}
\newunicodechar{λ}{\lambda}
\newunicodechar{Λ}{\Lambda}
\newunicodechar{μ}{\mu}
\newunicodechar{ν}{\nu}
\newunicodechar{ξ}{\xi}
\newunicodechar{Ξ}{\Xi}
\newunicodechar{π}{\pi}
\newunicodechar{Π}{\Pi}
\newunicodechar{ρ}{\rho}
\newunicodechar{σ}{\sigma}
\newunicodechar{Σ}{\Sigma}
\newunicodechar{τ}{\tau}
\newunicodechar{υ}{\upsilon}
\newunicodechar{ϒ}{\Upsilon}
\newunicodechar{φ}{\varphi}
\newunicodechar{ϕ}{\phi}
\newunicodechar{Φ}{\Phi}
\newunicodechar{χ}{\chi}
\newunicodechar{ψ}{\psi}
\newunicodechar{Ψ}{\Psi}
\newunicodechar{ω}{\omega}
\newunicodechar{Ω}{\Omega}

\def\tightmatrix{%
    \setcounter{MaxMatrixCols}{15}%
    \setlength{\arraycolsep}{1.2pt}%
    \renewcommand{\arraystretch}{0.6}%
}

\def\makecdotgray{%
	\let\oldcdot\cdot%
    \def\cdot{\color{gray}\oldcdot}%
}

\usepackage[export]{adjustbox}
\usepackage{xcolor}
\usepackage{booktabs}
\usepackage{amsmath,amssymb,mathtools,amsthm}
\usepackage{bm}
\usepackage[capitalise,noabbrev]{cleveref}
\newtheorem{theorem}{Theorem}
\newtheorem{conjecture}[theorem]{Conjecture}
\crefname{digression}{Digression}{Digressions}
\usepackage{tikz}
\usetikzlibrary{calc,shapes.geometric,shapes.symbols}
\def\e{\mathrm{e}}
\def\Po{\mathrm{Po}}
\def\Bin{\mathrm{Bin}}
\def\cp{c^{△}}

\def\csk{c^{⬜}_k}
\def\cpk{c^{△}_{k}}

\def\ck{c^{*}_{k}}
\def\ckl{c^{*}_{k,ℓ}}
\let\oldtextbf=\textbf
\def\textbf#1{\oldtextbf{\color{black!65}\boldmath #1}}
\def\textbfNoMath#1{\oldtextbf{\color{black!65} #1}}

\def\credits#1{\ \hfill \color{gray}\footnotesize #1}
\def\mycaptionstyle{\footnotesize}
\def\mycaption#1{\text{\mycaptionstyle #1}}

\def\vertCredit#1{
    \rotatebox{90}{\scriptsize\color{gray}#1}%
}
\def\xkcdCredit#1{%
    \vertCredit{\begin{tabular}{l}%
    adapted from
    \foreach \num in {#1}{
        \href{http://xkcd.com/\num}{xkcd/\num}
    }\\by Randall Munroe\end{tabular}}%
}

\newcounter{digression}
\newenvironment{digression}[2][hbt]{%
    \refstepcounter{digression}
    \begin{figure}[#1]
        \centering
        \begin{tikzpicture}
            \node[draw,thick,rounded corners,fill=black!10,font=\small] (box) \bgroup
                \begin{minipage}{0.9\textwidth}
                    \emph{\textbf{Digression \thedigression}: #2}\par
                    \color{black!80}
}{
                \end{minipage}
            \egroup;
        \end{tikzpicture}
    \end{figure}    
}

\title{What if we tried Less Power? \\[3pt] 
       \sffamily \large Lessons from studying the power of choices\\[-3pt]in hashing-based data structures}
\author{Stefan Walzer}

\begin{document}

\maketitle
\begin{abstract}
    The celebrated \emph{power of two choices} paradigm underlies \emph{cuckoo hash tables} as follows:
    If you have $n$ balls and $m = (2+ε)n$ bins and throw each ball into a bin at random, then likely some bin will receive $Ω(\frac{\log n}{\log \log n})$ balls. If, however, you can choose between \emph{two} random bins for each ball, you can likely arrange for a \emph{private bin} for each ball.
    
    In the first part of this article, we review some related space-efficient data structures on a high level.
    We'll find that the additional power afforded by \emph{more than $2$ choices} is often outweighed by the additional costs they bring.
    In the second part, we present a data structure where choices play a role at coarser than per-ball granularity. In some sense, we rely on the \emph{power of $1+ε$ choices} per ball.
\end{abstract}


{
    \noindent\itshape\footnotesize
    This article was written for the algorithms column of the bulletin of the EATCS. It is a “best-of” of my dissertation and related work reviewed from a fresh perspective. I've tried to make it a pleasant read conveying intuition while being unencumbered by technical details. So allow me to be your guide through the garden of my interests, present and past. Our winding path will circle a recent construction called \emph{Bumped Ribbon Retrieval} \cite{DHSW:Ribbon-SEA:2022}, with which our tour will conclude. 
}


\stepcounter{section}
\addcontentsline{toc}{section}{Data Structures using the Power of Two Choices}
\section*{Part 1: Data Structures using the Power of Two Choices}

The classical demonstration of the \emph{power of two choices} goes as follows \cite{ABKU:Balanced:1999,M:The_Power:1991}. Assume you have $n$ balls, $m = Θ(n)$ bins and you distribute the balls independently and uniformly at random into the bins.

\begin{center}
    \includegraphics[page=1]{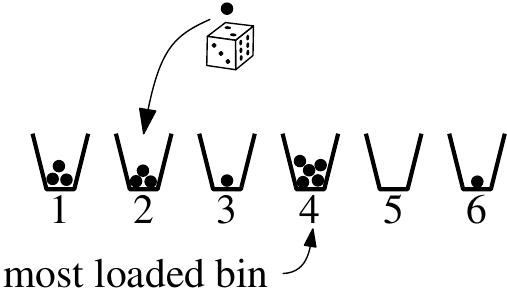}
\end{center}
Then the most loaded bin will contain $Θ(\frac{\log n}{\log \log n})$ balls with high probability (whp)%
\footnote{Defined as probability $1-o(1)$.}.
In contrast, assume you generate \textbf{two options} for each of the balls and place the balls sequentially, putting each ball into the \textbf{bin with the least load} among its two options (breaking ties arbitrarily).
\[
    \includegraphics[page=2]{img/ipe-pics.pdf}
\]
Now the \textbf{maximum load} is only $Θ(\log \log n)$ whp, which is \textbf{exponentially less}! The described setting is known as \textbf{online load balancing}. Bins might correspond to servers and balls to jobs that have to be assigned to servers on creation. What if we generate further options for each ball? For $d ≥ 2$ choices per ball, the maximum load becomes $\frac{\ln \ln n}{\ln d} + Θ(1)$ whp, i.e.\ $d$ only affects a constant factor. In other words, there is a massive difference between one and two choices and just a small difference between $2$ and $d ≥ 3$ choices. Hence the name power of \emph{two} choices.

Since its discovery, the power of two choices paradigm has been influential in data structure design, which is the focus of this paper.

\subsection{The Dictionary Problem \& Cuckoo Hashing}

Now consider \textbf{offline load balancing} where we generate the two random bins for all balls in advance and think carefully about all choices at the same time.
\begin{gather*}
    \includegraphics[page=3,width=0.9\textwidth]{img/ipe-pics.pdf}\\\notag
    \mycaption{Which option should each ball take to avoid collisions?}
\end{gather*}
It is helpful to use a different visualisation where each bin is a vertex and each ball an edge connecting the two bins it may be placed into:
\begin{equation}
    \includegraphics[page=4,width=0.75\textwidth]{img/ipe-pics.pdf}
    \phantom{m = 20}
    \label{pic:cuckoo-graph}
\end{equation}
If we ignore the possibility of duplicate edges, then we get a graph with $m$ vertices and $n$ uniformly random edges.%
This is known as an \emph{Erdős-Renyi random graph}. In their ancient and seminal paper “on the evolution of random graphs” \cite{ER:On_the_evolution:1960} Erdős and Renyi show that if the edge density satisfies $\frac{n}{m} < \frac{1}{2}-ε$ then whp all connected components of the graph are small \textbf{trees} ($\# \text{edges} = \# \text{vertices} - 1$) or pseudotrees ($\# \text{edges} = \# \text{vertices}$). For $\frac{n}{m} > \frac 12 + ε$, on the other hand, there is whp a “\textbf{giant component}” (comprising $Θ(n)$ vertices) that has more edges than vertices. Redrawing our example we find:
\begin{center}
    \includegraphics[page=5]{img/ipe-pics.pdf}
\end{center}
The task of placing all balls without collision corresponds to the task of assigning a \textbf{direction} to each of the edges in the graph such that every vertex has indegree at most $1$. For tree and pseudotree components this is easy: For trees, we pick an arbitrary root vertex and direct each edge away from the root. For a pseudotree, which contains a single cycle, we direct the cycle in some consistent way (say “clockwise”) and every other edge away from the cycle. For the remaining component(s) there is no solution by the pigeon hole principle.
\begin{center}
    \includegraphics[page=6]{img/ipe-pics.pdf}
\end{center}
The important observation here is: If we have $n$ balls and $m = (2+ε)n$ bins (for constant $ε > 0$) then \textbf{only trees and pseudotrees} arise whp and we can place all balls without a single collision.

\paragraph{Classic Cuckoo Hash Table.} This observation gives us a cuckoo hash table \cite{PR:Cuckoo:2004}, a simple data structure that stores $n$ elements, which we call \emph{keys}\footnote{In general elements could be key-value pairs, but values play no role in the following.}, using one\footnote{Most implementations use two arrays for reasons that need not concern us here.} array of $m$ memory cells and two hash functions that assign two uniformly random\footnote{See \cref{dig:suha}.} cells to each key.

\begin{digression}[tb]{Simple Uniform Hashing Assumption (SUHA).}
    \label{dig:suha}
    We assume that hash functions assign hash values to the keys independently. This \emph{simple uniform hashing assumption} is \textbf{unrealistic} as $\tilde{Ω}(n)$ bits of entropy would be needed for independence while popular practical hash functions like \emph{MurmurHash} \cite{Appleby:MurmurHash3:2012} or \emph{xxhash} \cite{Collet:xxhash:2020} use seeds of only $\tilde{𝒪}(1)$ bits.
    
    There are many standard ways of addressing this missmatch. We can try to work with weaker notions like \textbf{$k$-independence}, where any set of $k$ keys have independent hash values but any $k+1$ hash values may be correlated \cite{WC:New_Hash:1981}. A good overview on how this can help is given in \cite{Rink:Thesis:2015} and highly practical $2$-independent families are described in \cite{Thorup:High-Speed-Hashing:15}.
    A cryptographer might instead offer some insight into how \textbf{pseudorandomness} can be indistinguishable from randomness. \cite{AB:SipHash:2012}
    
    A well-subscribed lazy approach, that we also adopt here, is to simply use the SUHA as a \textbf{modelling assumption} and point to its excellent track record of capturing how popular hash functions behave in practice.
\end{digression}

Say we wish to store the set $\{☆,△,⬜,◯\}$.%
\footnote{We'll use comically undersized examples throughout this text that hopefully still get the point across. Practically relevant instances would typically have thousands or millions of keys.}
We consider the keys' hashes and find a \textbf{collision-free placement} that puts each key into one of its two assigned cells. (If no placement exists, we restart the construction with fresh hash functions.) To answer a query – say we wish to know if $△$ is in the set – we evaluate the hash functions on $△$ and search both cells for the requested key.
\begin{gather}
    \begin{tabular}{cc}
          \includegraphics[page=1]{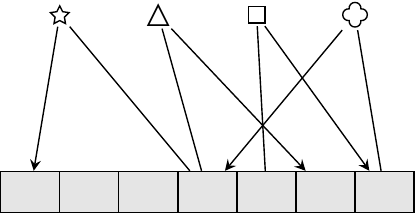}
        & \includegraphics[page=2]{img/table-pics.pdf}\\
          \mycaption{situation during construction}
        & \mycaption{situation during query}
    \end{tabular}
    \label{pic:bipartite-illustrations}
\end{gather}
\begin{minipage}{0.65\textwidth}
    Query times are \textbf{$𝒪(1)$} in the \textbf{worst-case} (not just in expectation). A down-side is that the \textbf{load factor} is $c = \frac{n}{m} < \frac{1}{2}$, meaning twice as much memory is required compared to naively storing keys consecutively. That's less than ideal. But what if we tried more power?
\end{minipage}\phantom{aa}
\begin{minipage}{0.25\textwidth}
    \includegraphics[width=2.5cm]{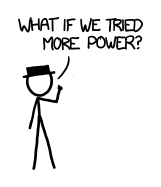}~
    \smash{\vertCredit{%
        \begin{tabular}{l}
            Illustration from 
            \emph{What If?}\\
            by Randall Munroe
        \end{tabular}%
    }}
\end{minipage}

\subsubsection{Higher Power: Generalisations of Cuckoo Hashing} 

A natural generalisation is to \textbf{assign more cells to each key} \cite{FPSS:Space_Efficient:2005}. For $k > 2$ cells the graph with $m$ vertices and $n$ edges from (\ref{pic:cuckoo-graph}) becomes a \emph{hyper}graph with $m$ vertices and $n$ \emph{hyper}edges of size $k$. This would make for a messy picture so we stick with bipartite illustrations like in (\ref{pic:bipartite-illustrations}).
\begin{equation}
    \raisebox{-1cm}{\includegraphics[page=3]{img/table-pics.pdf}}
    \label{fig:3-ary-cuckoo}
\end{equation}
Using $k > 2$ allows for the assumption $\frac{n}{m} < \frac{1}{2} - ε$ on the load factor to be relaxed. For $k = 3$ for instance, we find an increased \emph{load threshold} of $c₃^* ≈ 0.92$ up to which all keys can be placed whp. See \cref{dig:thresholds} for some background on the phenomenon of thresholds.

\begin{digression}[bt]{Sharp Load Thresholds.}
    \label{dig:thresholds}
    Consider the probability $p$ that all $n$ keys can be placed into a table of size $m$ when each key is assigned $k = 3$ cells at random. In the picture on the left, we plot (experimental approximations of) $p = p_m(c)$ for varying load factor $c = \frac{n}{m} ∈ [0,1]$ and fixed table size $m = 10⁴$.
    \begin{center}
        \includegraphics[page=1]{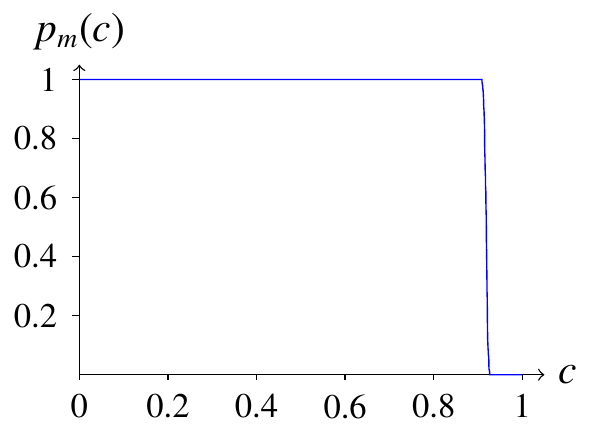}~
        \includegraphics[page=2]{exp/thresholds.pdf}
    \end{center}
    As expected, $p$ decreases as $c$ increases. What is \emph{not} obvious is that the transition from $p_m(c) ≈ 1$ to $p_m(c) ≈ 0$ happens within a tiny interval. On the right, we zoom in and also plot the function for $m = 10⁵$ and $m = 10⁶$. This shows that the \textbf{transition becomes steeper for larger $m$} and starts to resemble a \textbf{step function}. In fact, there is a \textbf{sharp load threshold} $c^*₃ ≈ 0.9179$ such that for $c < c^*₃$ we have $\lim_{m→∞}p_{m}(c) = 1$ and for $c > c^*₃$ we have $\lim_{m→∞}p_{m}(c) = 0$.
\end{digression}

Load thresholds $\ck$ are known for any $k$. They can be characterised implicitly as solutions to certain equations, but for our purposes, tabulated values will do. We include $c^*₁ = 0$ to emphasise that avoiding collisions with just \textbf{one hash function requires $m = Ω(n²)$} due to the \textbf{birthday paradox}, which gives a load factor of $\frac{n}{n²} → 0$.
\begin{center}    
    \renewcommand{\tabcolsep}{0.15cm}
    \begin{tabular}{c@{\qquad}ccccccc}
        \toprule 
        $k$ & 1 & 2 & 3 & 4 & 5 & 6 & 7\\
        \midrule
        $\ck$ &\phantom{0}0\phantom{0}&\phantom{00}0.5\phantom{00}& 0.91794 & 0.97677 & 0.99244 &  0.99738 & 0.99906\\
        \bottomrule
    \end{tabular}\nopagebreak\\[3pt]
    \credits{Values as determined in \cite{PR:Cuckoo:2004,DGMMPR:Tight:2010,FM:Maximum:2012,FP:Sharp:2012}.}
\end{center}
The values are of order $c_k^* = 1-\e^{-Θ(k)}$ (this can be derived form \cite{FP:Sharp:2012}) so we can achieve load factors arbitrarily close to $1$ by choosing $k$ large enough. However, any practitioner will be quick to point out that a query operation that has to check $k$ random memory locations is likely to incur \textbf{$k$ cache misses}, so increasing $k$ is not particularly enticing.

The far more popular generalisation sticks with $k = 2$ hash functions, but each hash value identifies a \emph{bucket} of $ℓ$ memory cells \cite{CSW:The_Random:2007,FR:The_k-orientability:2007,FKP:The_Multiple:2011}. Each key may then be placed into any cell in any one of its \textbf{two buckets}.
\begin{center}
    \includegraphics[page=4]{img/table-pics.pdf}
\end{center}
Thresholds $c_{2,ℓ}^*$ for the load factor achievable this way are also known:
\begin{center}    
    \renewcommand{\tabcolsep}{0.15cm}
    \begin{tabular}{cccccccc}
        \toprule 
        $ℓ$ & 1 & 2 & 3 & 4 & 5 & 6\\
        \midrule
        $c_{2,ℓ}^*$ &\phantom{a} \quad 0.5\quad \phantom{a}
        & 0.89701
        & 0.95915
        & 0.98037
        & 0.98955
        & 0.99407\\
        \bottomrule
    \end{tabular}\nopagebreak\\[3pt]
    \credits{Values as determined in \cite{PR:Cuckoo:2004,CSW:The_Random:2007,FR:The_k-orientability:2007,DGMMPR:Tight:2010,FKP:The_Multiple:2011}.}
\end{center}
Unsurprisingly, both avenues for generalisation can be combined such that a key can be placed within any of $k$ buckets of size $ℓ$ each. The corresponding thresholds $\ckl$ are all known \cite{FKP:The_Multiple:2011,L:A_New_Approach:2012,L:Belief_Propagation:2013}.

\paragraph{Which generalisation is better?}
Should we use \textbf{more hash functions or bigger buckets?} \emph{On the one hand}, if you consider the \textbf{number of memory cells a query touches} in the worst case, then using more hash functions seems superior to using bigger buckets. For instance, with $3$ hash functions we get a threshold of $c_{3}^* ≈ 0.918$ and have to scan three memory cells per query. When using $2$ hash functions and buckets of size $2$ we get a lower threshold of $c_{2,2}^* ≈ 0.897$ and have to scan four memory cells per query. It turns out the four correlated cells in the two buckets do not constitute as powerful a choice as three independent locations.

\emph{On the other hand}, the reduced number of \textbf{hash function evaluations and cache misses} strongly favours using larger buckets rather than additional hash functions, even if the number of memory cells associated with a key has to be higher to achieve the same load factor.

From what I have seen, practitioners aiming for high load factors seem to be pretty happy with either $k = 2$ or a middle ground using $k = 3$ hash functions and some bucket size $ℓ ∈ \{3,…,8\}$, see e.g.\ \cite{MS:DySECT-Journal:2017,ZR:cuckoo-for-k-mers:2022}. A related compromise assigns to each key $k$ cells \emph{within the same memory page} and $1$ additional cell in a \emph{backup page} \cite{DMR:Cuckoo:2011} (see also \cite{PS:A_Cuckoo_Hashing:2012}). It turns out most keys can then be stored on their primary page.

The message so far can be summarised as follows: 
Cuckoo hash tables are powered by the first two independent choices.
Adding the third choice can help but
competes for attention with other measures for increasing the load factor. \textbf{Two choices is all you really need.}

\subsubsection{Power Dynamics: Cuckoo Table Construction and Insertion}

The thresholds mentioned so far only relate to whether or not a rule-conforming placement of all keys in the hash table \emph{exists}. But how can such a placement be \emph{found and maintained}? For simplicity, we focus on cuckoo hashing with $k$ hash functions and buckets of size $ℓ = 1$.

\paragraph{Table construction.} 
Constructing a cuckoo hash table is about finding a matching in a bipartite graph such as (\ref{fig:3-ary-cuckoo}).
Out of the box, the maximum matching algorithm by Hopcroft and Carp \cite{HK:An_Algorithm:1973} has a \emph{worst case} running time of $𝒪(n^{3/2})$, but a specialised algorithm with \emph{expected} running time $𝒪(n)$ for $k ≥ 2$ is known \cite{Khosla:Balls-Into-Bins:2013}.

A conceptually interesting greedy algorithm is \emph{\textbf{peeling}}. 
It \textbf{identifies cells in the table that are an option for only one (remaining) key}. In the following illustration on the left, at first only $⬜$ can be placed in this way into cell $5$, then $☆$ and $◯$ can be placed, and, finally, $△$ ends up with three cells to itself and can be placed in any one of them.
\def\tikzTo#1{\tikz \draw[->,>=stealth] (0,0) -- (#1,0);}
\def\raisedPic#1{\raisebox{-1cm}{\includegraphics[page=#1]{img/table-pics.pdf}}}
\begin{center}
     \raisedPic{9}~\!\!\!\tikzTo{0.5}\!\!\!~
    \raisedPic{10}~\!\!\!\tikzTo{0.5}\!\!\!~
    \raisedPic{11}~\!\!\!\tikzTo{0.5}\!\!\!~
    \raisedPic{12}\\[2pt]
    \mycaption{The peeling process}
\end{center}
To better assess the utility of peeling, consider the load thresholds $\cpk$ up to which peeling manages to place all keys in a cuckoo hash table with $k$ hash functions:
\begin{center}    
    \renewcommand{\tabcolsep}{0.15cm}
    \begin{tabular}{cccccccc}
        \toprule 
        k & 3 & 4 & 5 & 6 & 7\\
        \midrule
        $\cpk$
        & 0.81847
        & 0.77228
        & 0.70178
        & 0.63708
        & 0.58178\\
        \bottomrule
    \end{tabular}\nopagebreak\\[3pt]
    \credits{Values as determined in \cite{Molloy05:Cores-in-random-hypergraphs}. Variants in \cite{C:Cores:2004,PSW:SuddenCore:96,Luczak:A-simple-solution} and \cite[Chapter 18]{MezMont:InfPhysComp:2009}.}
\end{center}
There are two things to notice here.
\emph{Less relevant} is that the value for $k = 2$ is missing. To see why, recall the graph with $m$ vertices and $n$ edges from (\ref{pic:cuckoo-graph}). Peeling can handle trees but gets stuck if there is at least one cycle. Unfortunately, the probability for a cycle to exist is bounded away from zero as soon as $\frac{n}{m}$ is bounded away from $0$. So for peeling to work \emph{whp} we need $k ≥ 3$.
\begin{minipage}{0.65\textwidth}
    \emph{More relevant} is that $\ck < \cpk$, meaning there are load factors where a \textbf{placement exists whp but peeling fails to find one whp}.
    Even worse, the gap $\ck - \cpk$ between the thresholds increases with $k$, even converging to $1$ for $k → ∞$.
    This \textbf{disqualifies peeling as a general-purpose construction} algorithm for cuckoo hash tables. However, peeling will make a comeback in more specific settings (stay tuned!).
\end{minipage}\phantom{aa}
\begin{minipage}{0.3\textwidth}
    \centering
    \raisedPic{16}\\[2pt]
    \mycaptionstyle{A placement exists, but peeling is stuck immediately.}
\end{minipage}

\paragraph{Insertions.}
A construction algorithm suffices for a \emph{static} key set, but for a dynamic data structure, we need to maintain a placement as keys are inserted and deleted over time.
A deletion is trivial: Simply locate the key, by checking all associated cells, and remove it from there. An insertion on the other hand amounts to modifying an existing matching to incorporate a new key. This suggests that we look for an \textbf{augmenting path}.
\begin{center}
    \raisedPic{18}
    \tikzTo{0.5}
    \raisedPic{19}
\end{center}
In the picture, $☆$ is the newly added key and moves into the cell previously used by $⬜$, which moves into the cell previously used by $◯$, which moves into an empty cell.

There are two well-known strategies for finding such an augmenting path, both proposed in \cite{FPSS:Space_Efficient:2005}. \emph{\textbf{Breadth first search} (BFS)} insertion computes the \textbf{shortest augmenting path} in the natural BFS way. \textbf{\emph{Random walk} (RW)} insertion goes ahead and places the unplaced key into one of its cells at random and \emph{evicts} the key that was there before (if any). The evicted key is then also placed randomly and so on until an evicted key is placed into an empty cell. Strategies more clever than this have also been considered, some of which store some auxiliary data in the cells \cite{K:Fast-Kickout-Schemes:2016}. How good are these algorithms? It seems that, up to constant factors between them, they are equally \textbf{excellent in practice} in the sense that their running time does not depend on $n$, i.e.\ is $𝒪(1)$.
\begin{conjecture}[{Echoing sentiments from
 \cite{FJ:Insertion-time-Cuckoo-journal:2019,W:Cuckoo-RW-Insertion:2022,K:Fast-Kickout-Schemes:2016,FMM:An_Analysis:2011,FPSS:Space_Efficient:2005}}]
    Consider a cuckoo hash table with $k ≥ 2$ hash functions, buckets of size $ℓ ≥ 1$, and a load factor $\frac{n}{m} < \ckl - ε$ for some $ε > 0$. Assume the keys are inserted sequentially using BFS insertion or RW insertion. Conditioned on the high probability event that a placement of all keys exists, the expected time to perform each insertion is $𝒪(f(ε))$ for some function $f$ that does not depend on $n$.
\end{conjecture}
I know of \textbf{no proof}, neither for RW nor for BFS and for no pair of $k$ and $ℓ$, except for the classical case of $k = 2$ and $ℓ = 1$,
where there is no choice regarding which key to evict from a given bucket (there is just one) and no choice regarding where to relocate an evicted key to (there is just one alternative).
%
Partial proofs exist for the case with $ℓ = 1$ for
\begin{itemize}
    • BFS, for $k > 8$ and under a stronger restriction on the load factor \cite{FPSS:Space_Efficient:2005},
    • RW, for large $k$  and under a stronger restriction on the load factor \cite{FJ:Insertion-time-Cuckoo-journal:2019},
    • RW, for load factors below the \emph{peeling} threshold $\cpk$ \cite{W:Cuckoo-RW-Insertion:2022},
    • RW, but only guaranteeing running times of $𝒪(\mathop{\mathrm{polylog}} n)$ \cite{FMM:An_Analysis:2011,FPS:On_the_Insertion:2013}.
\end{itemize}
More progress towards proving the conjecture would be exciting. Maybe techniques from statistical physics, which have been a powerful tool for determining thresholds in the static case \cite{L:A_New_Approach:2012,Leconte:Cuckoo:2013,L:Belief_Propagation:2013} can help with the dynamic case as well.

\subsubsection{Creatively Wielding the Power}
\label{sec:hashing-scheme-gallery}

The number $k$ of hash functions and the size $ℓ$ of buckets are not the only degrees of freedom in the design space. Here is a short list of \textbf{further variants} that were considered and the reasons why.
\begin{gather}
    \label{pic:cuckoo-variants}
    \ \hspace{-5em}\begin{tabular}{ccc}
         \mycaption{\textbf{Double Hashing} \cite{MT:Peeling_Arguments:2012}}
        &\mycaption{\textbf{Unaligned Blocks} \cite{LP:3.5-Way:2009}}
        &\mycaption{\textbf{Spatial Coupling} \cite{W:SpatialCoupling:2021}}
        \\
         \includegraphics[page=6,valign=t]{img/table-pics.pdf}
        &\includegraphics[page=7,valign=t]{img/table-pics.pdf}
        &\includegraphics[page=8,valign=t]{img/table-pics.pdf}
    \end{tabular}\hspace{-5em}
\end{gather}
\begin{description}
    •[Double Hashing.] Mitzenmacher and Thaler \cite{MT:Peeling_Arguments:2012} proposed that the buckets $b₁,…,b_k$ assigned to a key are not chosen independently. Instead, only $b₁$ and an offset $d$ are chosen at random, and $b₂,…,b_n$ are defined as $b_i := b₁ + (i-1)·d$, modulo the number of buckets. This \textbf{reduces} the amount of \textbf{entropy} in a key's hash values from $k\log m$ to $2\log m$ with no apparent downsides. In particular the thresholds $\ckl$ remain the same \cite{Leconte:Cuckoo:2013,MPW:DoubleHashing:2018}.
    •[Unaligned Blocks.] Lehman and Panigrahy proposed to use buckets that do not form a partition of the set of cells. Rather, \emph{any contiguous sequence} of $ℓ$ cells can occur as a bucket, regardless of the offset. This scheme yields \textbf{higher thresholds} than $\ckl$ without affecting the access pattern of queries \cite{LP:3.5-Way:2009,Walzer:OverlappingBlocks:2018}.
    •[Spatial Coupling.] Walzer proposed that the $k$ buckets assigned to a key are chosen within the same interval of $εm$ buckets for some $ε > 0$. Assuming $ε$ is small enough and the load factor is less than $\ckl$, not only does a placement exist whp, but \textbf{peeling works whp}. \cite{W:SpatialCoupling:2021}
\end{description}
I cannot help but wonder which other surprising effects can be achieved by further creative cuckoo hashing variants.

\subsection{The Retrieval Problem \& Random Matrices}

In the \emph{\textbf{retrieval problem}}\footnote{The first mention of the problem under the name “retrieval” that I could find is in \cite{DMPP:De_Dictionariis:2006} in 2006. Bloomier filters \cite{CKRT:The_Bloomier:2004} in 2004 are a clear spiritual predecessor and related to approximate membership. More background is given in \cite{DP:Succinct:2008}.} we are given a set $S$ and a function $f : S → \{0,1\}^r$ for some $r ∈ ℕ$. Let us assume $r = 1$. The task is to construct a data structure that returns $f(x)$ when queried for $x ∈ S$. What is unusual is that a query for some \textbf{$y ∉ S$} may return an \textbf{arbitrary result}. In an instructive example due to Pagh and Dietzfelbinger \cite{DP:Succinct:2008}, $S$ is a set of names and $f$ tells us if a name $x ∈ S$ is female ($f(x) = 1$) or male ($f(x) = 0$).
\begin{center}
    \includegraphics[page=7]{img/ipe-pics.pdf}
\end{center}

When $f$ reflects typical English names, the data structure should return $0$ for $\textsc{john} ∈ \mathrm{domain}(f)$ and $1$ for $\textsc{mary} ∈ \mathrm{domain}(f)$. When queried for $\textsc{banana} ∉ \mathrm{domain}(f)$ both $0$ and $1$ are allowed, we don't care.

The \textbf{trivial solution} for this problem stores the \textbf{set of pairs}
\[ f = \{(\textsc{john},0),(\textsc{mary},1),(\textsc{lizzy},1),(\textsc{paul},0),…\}\]
using a dictionary. This requires storing at least $n = |S|$ strings. However, we will soon see that the most space-efficient solutions require little more than $n$ \emph{bits}. Note that $n$ bits are necessary if we make no further assumptions on the input.\footnote{If there are regularities such as the majority of names in $\mathrm{domain}(f)$ being male or most female names ending in a vowel, then compression may be possible, see \cite{HKP:CompressedFunction:2009,BelazzouguiV13,GOV:retrieval-Compressed:2020}.}

As a warm-up, here is a \textbf{compact solution} that needs $𝒪(n)$ bits. We use an array of $m = 𝒪(n)$ cells that can store values from $\{0,1,⊥\}$. We associate each $x ∈ S$ with a random cell $h(x) ∈ [m]$ and set a cell to $0$ or $1$ whenever a consistent value exists and $⊥$ in case of conflicts:
\begin{center}
    \includegraphics[page=13]{img/table-pics.pdf}
\end{center}
The keys involved in a conflict make up a constant fraction of all keys in expectation. For these, we can build a fallback data structure recursively. Some readers may have fun working out that we get \textbf{constant expected access time} and a total space consumption of roughly $\e/2$ array cells per key for $m = n/2$.%
\footnote{Hint: Assume that $n/2$ names are male and $n/2$ names are female (you can later check that this is the worst case). Then use that for a fixed name $x$, the number of names of the opposite gender that share the hash value of $x$ has distribution $\Bin(\frac n2, \frac 1m)$, which converges to the Poisson distribution $\Po(\frac n{2m}) = \Po(1)$ that satisfies $\Pr[\Po(1) > 0] = 1-1/\e$.}
This amounts to \textbf{$\e ≈ 2.72$ bits per key} when using the naive encoding $\{0 ↦ 00, 1 ↦ 11, ⊥ ↦ 01\}$.%
An improved version of this idea is known as \textbf{filtered retrieval} \cite{Sanders:Retrieval-FingerPrinting:2014}.

\subsubsection{The Power to be Independent: Retrieval via Random Linear Systems}

To get closer to \textbf{succint retrieval} we consider strange cousins of cuckoo hash tables, cf.~\cite{BPZ:Practical:2013,GL:XorFilters:2020}. While not employing the power of two choices in the traditional sense, their setup and analysis are closely related. Hash functions assign to each key \textbf{several cells} in an array of size $m ≥ n$ that is populated with bits in such a way that taking the \textbf{\textsc{xor} of all bits} associated with $x ∈ S$ yields $f(x)$.
\begin{equation}
    \includegraphics[page=14]{img/table-pics.pdf}
    \label{pic:retrieval}
\end{equation}
A query for \textsc{john} would, for instance, compute $f(\textsc{john}) = 0⊕1⊕1 = 0$, where $⊕$ denotes \textsc{xor}. To construct the data structure we had to choose values $z₁,z₂,…,z₇ ∈ \{0,1\}$ to put into the array to simultaneously satisfy the equations
\begin{align*}
    \begin{matrix}
    z₁ ⊕ z₂ ⊕ z₄ = 0\\
    z₂ ⊕ z₄ ⊕ z₆ = 1\\
    z₁ ⊕ z₆ ⊕ z₇ = 0\\
    z₄ ⊕ z₆ ⊕ z₇ = 1
    \end{matrix}
    \color{black!60}
    \qquad
    \begin{matrix}
         \textsc{(john)}\\
         \textsc{(mary)}\\
         \textsc{(paul)}\\
         \textsc{(lisa)}
    \end{matrix}
\end{align*}
Since $⊕$ is addition in the \textbf{two element field} $𝔽₂ = \{0,1\}$ these equations are \textbf{linear equations} over $𝔽₂$.
So fasten your seatbelt for a bit of linear algebra, but don't worry, it's not so bad.
We can write the above equations in \textbf{matrix form} as
\[
    \begin{pmatrix}
        1 & 1 & 0 & 1 & 0 & 0 & 0\\
        0 & 1 & 0 & 1 & 0 & 1 & 0\\
        1 & 0 & 0 & 0 & 0 & 1 & 1\\
        0 & 0 & 0 & 1 & 0 & 1 & 1
    \end{pmatrix}
    · 
    \begin{pmatrix}
        z₁\\
        z₂\\
        z₃\\
        z₄\\
        z₅\\
        z₆\\
        z₇
    \end{pmatrix}
    = 
    \begin{pmatrix}
        0\\
        1\\
        0\\
        1
    \end{pmatrix}\quad
    \color{black!60}
    \begin{matrix}
        \textsc{(john)}\\
        \textsc{(mary)}\\
        \textsc{(paul)}\\
        \textsc{(lisa)}\\
    \end{matrix}
\]
When does such a system have a solution $\vec{z} ∈ \{0,1\}^m$? Well, the \emph{columns} of the $n × m$ matrix $A$ should better span all of $\{0,1\}^n$, so that the right hand side vector $(f(x))_{x ∈ S} ∈ \{0,1\}^n$ can surely be attained as a linear combination of these columns. In other words, the column rank of $A$ should be $n$. Since column rank and row rank are the same thing, the $n$ \emph{rows} of $A$ have to be \textbf{linearly independent}.

\paragraph{What are our goals here?} Remember that $n$ is part of the input and we are free to choose two things: The number $m$ of columns and the way in which keys are associated with row vectors via hash functions. Ideally we want that \textbf{$m$ is small} so that $z ∈ \{0,1\}^m$ is \textbf{cheap to store} and we want row vectors to have \textbf{small Hamming weight} so queries are \textbf{cheap to evaluate}. Both of these goals are intuitively in tension with the independence requirement.

An encouraging fact is that even \emph{square} matrices (i.e.\ $m = n$) where every entry is chosen by a biased coin with probability $p = \frac{\log n}{n}$ (i.e.\ rows have expected Hamming weight $\log n$) are regular with constant probability \cite{Cooper:Rank-Of-Random-Matrices:2000}. This is, however, not the most natural setup for our purposes.

\subsubsection{New Data Structure, Same Thresholds:
\texorpdfstring{\\}{} How Cuckoo Hashing Connects to Retrieval}

A more natural setup already depicted in (\ref{pic:retrieval}) associates $k$ random cells with every key, which produces a matrix with \textbf{$k$ ones per row} in random positions.

There is a \textbf{threshold} $\csk$ for the ratio $c = \frac{n}{m}$ such that $A$ has rank $n$ whp when $c < \csk -ε$ and $A$ has rank less than $n$ whp when $c > \csk + ε$. Remarkably, it \textbf{coincides with the threshold for cuckoo hashing} with $k$ hash functions, i.e.\ $\csk = \ck$. Only the direction $\csk ≤ \ck$ is easy to show here. If $A$ has rank $n$, then some selection of $n$ columns induces a regular submatrix $A'$ (highlighted below).
\[
    \def\One{1}
    \def\Zer{0}
    \def\ONE{\bm{1}}
    \def\ZER{\bm{0}}
    \def\M#1#2{%
        \tikz[remember picture,inner sep=0,baseline=(pos#1.base)] \node (pos#1) {$#2$};%
    }
    \begin{pmatrix}
        \M1\One &    \ONE & \Zer & \M3\One & \Zer & \Zer & \M5\Zer\\
           \Zer &    \One & \Zer & \ONE    & \Zer & \One &    \Zer\\
           \ONE &    \Zer & \Zer & \Zer    & \Zer & \One &    \One\\
           \Zer & \M2\Zer & \Zer & \M4\One & \Zer & \One & \M6\ONE
    \end{pmatrix}
    \begin{tikzpicture}[overlay,remember picture]
        \foreach \i/\j in {1/2,3/4,5/6}{
            \fill[fill opacity=0.2,black] ($(pos\i.north west) + (-1pt,1pt)$) rectangle ($(pos\j.south east) + (1pt,-1pt)$);
        }
    \end{tikzpicture}
\]
The determinant of $A'$ is a non-zero number in $𝔽₂$, hence $\det(A') = 1$. By the Leibniz formula for determinants we have $1 = \det(A') = \sum_{π ∈ S_n} \prod_{i ∈ [n]} a'_{i,π(i)}$ where $(a'_{i,j})_{i,j∈[n]}$ are the entries of $A'$ and $S_n$ is the symmetric group with $n$ elements. At least one $π ∈ S_n$ must yield a non-zero contribution to $\det(A')$, and all entries $a'_{i,π(i)}$ for $i ∈ [n]$ must then be $1$. These entries are shown in bold in the matrix above and correspond to an injective placement of all keys in a cuckoo hashing setting. Bluntly:
\[ c < \csk ⇔ \text{“retrieval works whp”} ⇒ \text{“cuckoo hashing works whp”} ⇔ c < \ck\]
hence $\csk ≤ \ck$. The converse statement $\csk ≥ \ck$ requires a lot more work to prove \cite{DM:The_3-XORSAT:2002,PS:The_Satisfiability:2016,DGMMPR:Tight:2010,CKKR:k-XORSAT:2023}.

\subsubsection{Longer Blocks and Bitparallel Queries}

To get \textbf{thresholds closer to $\bm{1}$} there are similar options as in cuckoo hashing. \textbf{Increasing $k$} works but requires queries to combine data from more independent cells, resulting in \textbf{more cache misses}. Meh. We can also adopt the idea of using buckets of size $ℓ$. Naively connecting each key to each cell in two blocks of size $ℓ$ \emph{does not work} however.
We merely obtain copies of identical columns that do not contribute to the matrix rank, as shown below on the left (from now on we use a dot “$·$” to indicate implicit zeroes that do not explicitly occur in any representation and we omit the right hand side of equations). \emph{Instead} we should associate each key with a \textbf{random non-empty \emph{subset} of cells of each of its blocks} as shown on the right.
\begin{equation}
    \def\z{···}
    \setlength{\arraycolsep}{1.5pt}
    \begin{minipage}{5cm}
        {\centering%
        $\begin{pmatrix}
                    \z&111&\z&111&\z\\
                    111&\z&\z&\z&111\\
                    \z&111&\z&\z&111\\
                    111&\z&111&\z&\z\\
                    \z&111&111&\z&\z\\
                    111&\z&\z&111&\z\\
                \end{pmatrix}$

        }
        \mycaptionstyle \textbf{bad idea}: each key (row) is associated with all cells within two blocks of size $ℓ = 3$.
    \end{minipage}\hspace{1cm}
    \begin{minipage}{5cm}
        {\centering
        $\begin{pmatrix}
            \z&010&\z&110&\z\\
            100&\z&\z&\z&110\\
            \z&111&\z&\z&010\\
            010&\z&110&\z&\z\\
            \z&111&001&\z&\z\\
            101&\z&\z&001&\z\\
        \end{pmatrix}$

        }
        \mycaptionstyle \textbfNoMath{okay idea}: each key (row) is associated with \emph{some} cells within two blocks of size $ℓ = 3$.
    \end{minipage}
    \label{pic:two-block-matrix}
\end{equation}
Precise thresholds are not known and not identical to the cuckoo hashing thresholds $c^*_{2,ℓ}$. What we do know is that for suitable $ℓ = Θ(\log n)$ and $m = n + Θ(\log n)$ we obtain a matrix with rank $n$ whp  \cite{DW:Retrieval-log-extra-bits:2019}. This yields a \textbf{succinct} retrieval data structure with an almost optimal space requirement of $n + 𝒪(\log n)$ bits. Now consider a query. Given the solution vector $z ∈ \{0,1\}^m$ and the row vector associated with a key $x ∈ S$ we have to compute their \textbf{scalar product} to obtain $f(x)$.
\begin{center}
    \def\tikzmark#1#2{%
        \tikz[remember picture,inner sep=1,baseline=(#1.base)] \node (#1) {#2};%
    }
    \begin{tabular}{r@{\,}r@{\,}c@{\,}c@{\,}c@{\,}c@{\,}c@{\ }l@{}l}
        solution vector $z^T$: & (
        & {111} & {110} & {010} & {101} & {111}&)\tikzmark{l1}{\ }\\[-2pt]
        query(\textsc{jane}) $\xrightarrow{\text{hashing}}$ row vector for \textsc{jane}: & (
        & ${···}$ & {011} & ${···}$ & ${···}$ & {101} &)\tikzmark{l2}{\strut}\\
        \cline{2-8}
        & & & \tikzmark{v1}{010} &&& \tikzmark{v2}{101} &\tikzmark{l3}{\strut}\\
        &&&&&&&& $\tikzmark{pc}{popcount} = 3$\\
        &&&&&&&& \qquad $\tikzmark{ans}{1} \pmod 2$
    \end{tabular}%
    \begin{tikzpicture}[overlay,remember picture,>=stealth,semithick]
        \draw[->] (v1.south) 
        .. controls ($(v1.south)+(0,-0.3)$)
        and         ($(pc.west)+(-1,0)$) .. (pc.west);
        \draw[->] (v2.south) 
        .. controls ($(v2.south)+(0,-0.3)$)
        and         ($(pc.west)+(-0.5,0)$) .. (pc.west);
        \draw[->] (pc.210) to[bend right] (ans);
        \draw[->] (l2)
            .. controls ($(l2)+(0.5,0)$) and ($(l3)+(0.5,0)$) .. (l3)
            node[pos=0.5,right] {bitwise $\textsc{and}$};
    \end{tikzpicture}
\end{center}
\vspace{-0.5\baselineskip}
An element-wise product of two vectors in $𝔽₂^ℓ$ is a \textbf{bitwise \textsc{and}} and the sum of the entries of a vector in $𝔽₂^ℓ$ is a \textbf{population count modulo $2$}. On a \textbf{word RAM} with word size $w = Ω(ℓ)$, we can perform all we need for a \textbf{query in $𝒪(1)$} steps.

With constant time queries and almost optimal space, this seems like an ideal data structure until you realise that there is no fast way to construct it.

\subsubsection{To Gauss or not to Gauss: Constructing Retrieval Data Structures}

A problem with all this is that computing the solution vector $z ∈ \{0,1\}^m$ \textbf{requires solving a system} of linear equations. Spending $𝒪(n³)$ time on Gaussian elimination seems prohibitively expensive. Let us consider some alternatives.

\paragraph{Linear Time Solvers using Peeling.}
We can use a setup where peeling works whp, traditionally with $k = 3$ cells per key and a load factor below $\cp₃ ≈ 0.82$ \cite{BPZ:Practical:2013,Molloy05:Cores-in-random-hypergraphs}. Peeling means we look for a variable only appearing in one equation. We postpone initialising this variable to the end and ignore the equation until then. The remaining system may again have a variable appearing in only one of the remaining equations and so on. A different way of saying that a linear system is peelable is that its matrix can be transformed into echelon form by \textbf{row and column \emph{permutations} alone}, with no need for row \emph{additions}.
\begin{center}
    \def\to{\raisebox{1cm}{\!\!$→$}\!\!}
    $\displaystyle
       \includegraphics[page=1]{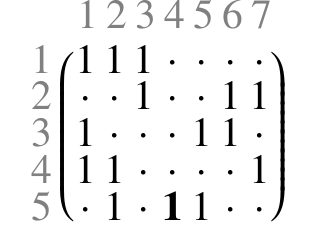}
    \to\includegraphics[page=2]{img/tightMatrix.pdf}
    \to\includegraphics[page=3]{img/tightMatrix.pdf}
    \to\includegraphics[page=4]{img/tightMatrix.pdf}$\\
    \mycaption{The row and column of the highlighted $1$ is swapped to the front to create an echelon form.}
\end{center}
After finding the peeling order of equations and variables in \textbf{linear time}, we can find a solution with back substitution in linear time. The setup with peeling thresholds close to $1$ from \cref{sec:hashing-scheme-gallery} reconciles this approach with close to optimal space efficiency both in theory \cite{W:SpatialCoupling:2021} and in practice \cite{ML:Fuse-Filters:2022}.

\paragraph{Quadratic Time Solver using Wiedemann's Algorithm.} If $𝔽$ is a finite field, $A ∈ 𝔽^{n × n}$ a regular matrix with $ψ$ non-zero entries and $b ∈ 𝔽^n$ then a solution $z$ to $A·z = b$ can be computed in $𝒪(nψ)$ field operations using \textbf{Wiedemann's algorithm} \cite{W:Solving:1986}. For sparse matrices with $ψ = 𝒪(n)$ this gives a \textbf{quadratic time} solver that can be generalised to non-square matrices $A$. Exploiting this for retrieval has been tried \cite{ADR:Experimental:2009,W:Thesis}, but it did not perform convincingly in experiments.

\paragraph{Bringing out the Big Gauss Rifles.}
A line of papers starting with Genuzio et al \cite{Vigna:Fast-Scalable-Construction-of-Functions:2016} cooked up ideas that made biting the bullet of performing Gaussian elimination seem like an almost appetising prospect.


The most important realisation is that we can \textbf{partition the input set} into many \emph{shards} of expected size $C$ using a hash function and construct a data structure for each of these shards. This reduces the running time from $𝒪(n³)$ to $𝒪(\frac{n}{C}·C³) = 𝒪(nC²)$. The price is an additional level of indirection during queries and a few bits of metadata for each of the $\frac{n}{C}$ shards.\footnote{In theory, a construction with two levels of indirection and shards of size $C = \sqrt{\log n}$ can lead to a succinct data structure with linear construction time \cite{P:An_Optimal:2009}. However, astronomical input sizes are required for the asymptotic behaviour to kick in.}
This can be combined with \textbf{bit-level parallelism}, a \textbf{structured Gaussian elimination} that tries to keep the matrix as sparse as possible for as long as possible, and the \textbf{method of four Russians} \cite{ADKF:FourRussians:1970}, which eliminates entries from $𝒪(\log n)$ columns at the same time. This yields fairly efficient solvers that are trivial to  \textbf{parallelise}.

\stepcounter{section}
\addcontentsline{toc}{section}{The Power of less than two Choices}
\section*{Part 2: The Power of \emph{less than two} Choices}
\subsection{Ribbon: Choice is overrated}
\label{sec:standard-ribbon}

After a talk on the retrieval data structure using two blocks per matrix row (see (\ref{pic:two-block-matrix}) on the right) a listener asked a question.
\begin{center}
    \begin{minipage}{0.28\textwidth}
        \xkcdCredit{2768}
        \includegraphics[width=0.5\textwidth]{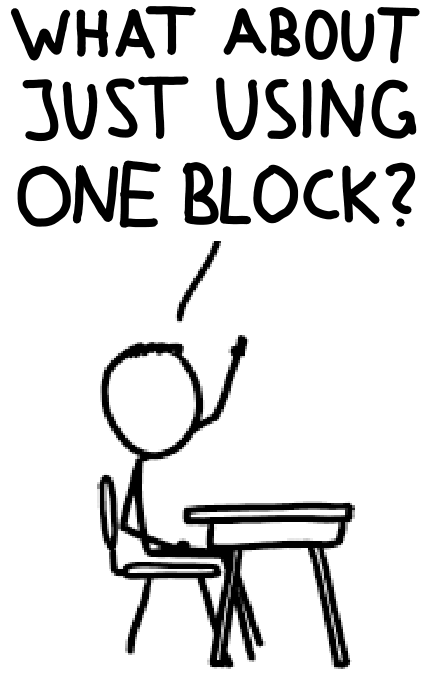}
        \nopagebreak\\
        \mycaptionstyle{Seth Pettie to my coauthor Martin Dietzfelbinger at Dagstuhl-Seminar 19051.}
    \end{minipage}\hspace{0.1\textwidth}
    \begin{minipage}{0.4\textwidth}
        Supposedly he had this in mind:
        \begin{equation}
            \makecdotgray
            \tightmatrix
            \begin{pmatrix}
                ·&·&·&·&·&·&1&1&1&1&·&·&·\\
                ·&·&·&·&0&0&1&1&·&·&·&·&·\\
                1&0&1&1&·&·&·&·&·&·&·&·&·\\
                ·&·&·&·&·&·&1&0&1&1&·&·&·\\
                ·&·&·&·&1&0&0&1&·&·&·&·&·\\
                ·&·&·&·&·&·&·&·&·&1&0&0&1\\
                ·&·&·&·&·&0&1&0&1&·&·&·&·\\
                ·&·&·&·&·&·&1&1&0&0&·&·&·\\
                ·&·&1&1&0&0&·&·&·&·&·&·&·\\
                ·&·&·&·&·&1&0&1&1&·&·&·&·
            \end{pmatrix}
            \label{eq:1-block-matrix}
        \end{equation}
    \end{minipage}
\end{center}
In the suggested scheme there is a block size $w$ and each key is associated with a \emph{starting position} $i ∈ [m-w+1]$ and a coefficient vector $c ∈ \{0,1\}^w$ that together lead to a row in $A$ with zeroes everywhere except for \textbf{one randomly placed block of $w$ random bits}.

This blasphemous single-block-per-key suggestion throws the power of choices paradigm completely out the window. Could this work?\footnote{Spoiler alert: Yes, and it kicked off an entire line of papers for us \cite{DW:One-Block-per-Row:2019,Ribbon:Arxiv:2021,DHSW:Ribbon-SEA:2022}.} More precisely, for which values of $m,n$ and $w$ is such a matrix likely to have independent rows?

\paragraph{Simplification: Bounded Linear Probing.}
First, let us \textbf{sort the rows} by starting position. Second, let us simplify the problem by ignoring the patterns of zeroes and ones and just \textbf{consider the placement problem} where we wish to select one position within the block of each row without selecting the same column twice.
Our question is now simply: In a linear probing hash table where each key hashes to a starting position, \textbf{can all keys be placed within $w-1$ cells of their starting positions?}
\begin{center}
    \begin{tabular}{ccc}%
        \makecdotgray%
        \tightmatrix%
        $\begin{pmatrix}
            1&0&1&1&·&·&·&·&·&·&·&·&·\\
            ·&·&1&1&0&0&·&·&·&·&·&·&·\\
            ·&·&·&·&1&0&0&1&·&·&·&·&·\\
            ·&·&·&·&0&0&1&1&·&·&·&·&·\\
            ·&·&·&·&·&1&0&1&1&·&·&·&·\\
            ·&·&·&·&·&0&1&0&1&·&·&·&·\\
            ·&·&·&·&·&·&1&0&1&1&·&·&·\\
            ·&·&·&·&·&·&1&1&0&0&·&·&·\\
            ·&·&·&·&·&·&1&1&1&1&·&·&·\\
            ·&·&·&·&·&·&·&·&·&1&0&0&1
        \end{pmatrix}$%
        &
        \def\item{\text{\textbullet}\vphantom{1}}
        \def\M#1{\tikz[overlay, remember picture] \coordinate (#1);}%
        \makecdotgray%
        \tightmatrix%
        $\begin{pmatrix}
            •&∘&∘&∘&\M{tl} ·&·&·&·&·&·&·&·&·\\
            ·&•&∘&∘&∘&·&·&·&·&·&·&·&·\\
      \M{TL}·&·&·&·&•&∘&∘&∘&·&·&·&·&·\\
            ·&·&·&·&∘&•&∘&∘&·&·&·&·&·\\
            ·&·&·&·&·&∘&•&∘&∘&·&·&·&·\\
            ·&·&·&·&·&∘&∘&•&∘&·&·&·&·\\
            ·&·&·&·&·&·&∘&∘&•&∘&·&·&·\\
            ·&·&·&·&·&·&∘&∘&∘&•&·&·&·\\
            ·&·&·&·&·&·&∘&∘&∘&∘&·&·&·\M{BR}\\
            ·&·&·&·&·&·&·&·&·&∘\M{br} &•&∘&∘
        \end{pmatrix}$
        \begin{tikzpicture}[overlay, remember picture]
            \fill[black,fill opacity=0.12] (tl) ++ (-3pt,10pt) rectangle ($(br)+(2pt,-2pt)$);
            \fill[black,fill opacity=0.12] (TL) ++ (-3pt,10pt) rectangle ($(BR)+(2pt,-2pt)$);
        \end{tikzpicture}
        &
        \raisebox{-1cm}{\includegraphics[page=15]{img/table-pics.pdf}}
        \\
         \begin{minipage}[t]{0.3\textwidth}\mycaptionstyle\centering
            rows sorted by\\starting position
         \end{minipage}
        &\begin{minipage}[t]{0.3\textwidth}\mycaptionstyle\centering
            simplified task: select one position in the
            block of each row using distinct columns
         \end{minipage}
        &\begin{minipage}[t]{0.3\textwidth}\mycaptionstyle\centering
            bounded linear probing
         \end{minipage}
    \end{tabular}
\end{center}
A greedy algorithm suffices to decide this question: Go through the keys in ascending order of starting position and place each key as far left as possible. The filled dots above indicate where the keys are placed. If we fail to place a key this way, as we do in the example, then this is witnessed by a range of $N$ positions such that at least $N+1$ keys have their block completely contained within the range (shaded grey with $N = 6$).

\paragraph{Connection to queuing theory.}
There is an equivalent way to describe the greedy placement algorithm just discussed (cf.\ \cite{DW:One-Block-per-Row:2019}).
We maintain a \textbf{FIFO queue} $Q$ of keys and go through the table cells from left to right. When handling cell $i$ we add the keys with starting position $i$ to the back of $Q$ and then place the first key in $Q$ (if any) into position $i$ and remove it from $Q$. This procedure places every key within its block if and only if the size of $Q$ never exceeds $w$.\footnote{Can you see why?} We therefore analyse the size $q_i$ of $Q$ after step $i$. If $x_i$ is the number of keys with starting position $i$ then
\begin{equation}
    q_i = \max(0, q_{i-1} + x_i - 1) \text{\ for $i ≥ 1$, and $q₀ = 0$.}
    \label{eq:m/d/1-queue}
\end{equation}
Each $x_i$ has Binomial distribution $\Bin(n,\frac{1}{m-w+1})$ because each of the $n$ keys has an independent chance of $\frac{1}{m-w+1}$ to have $i$ as its starting position. Since we know $\sum_i x_i = n$, there is a slight but annoying negative correlation between the $x_i$. A common approximation in such a situation \textbf{replaces} the family $(x_i)_{i ∈ [m-w+1]}$ of \textbf{Binomial} random variables with an \emph{independent} family $(x_i')_{i ∈ [m-w+1]}$ of \textbf{Poisson} distributed random variables with the same expectation $𝔼[x_i'] = α := \frac{n}{m-w+1}$. See \cref{dig:poissonisation} for an explanation.

\begin{digression}{Poissonisation and random coupling (see also \cite[Chapter 5.4]{MU:Probability:2017}).} 
    \label{dig:poissonisation}
    {\centering
        \includegraphics[page=8]{img/ipe-pics.pdf}\\
    }

    The picture shows a strip of width $m-w+1$ that is open to the top and contains points randomly with a \emph{rate} of $1$ point per unit of area (known as a \emph{Poisson point process}).
    Within the strip we consider $m-w+1$ disjoint sub-strips of height $α$ as shown. The number of points $x_i'$ within the $i$-th strip has distribution $\Po(α)$. Let now $H$ be the smallest vertical position such that exactly $n$ points fall at or below $H$. Note that $H$ is a random variable. Because each of the points below $H$ is within any of the vertical strips with the same probability, the number $x_i$ of points below $H$ in the $i$-th strip has distribution $\Bin(n,\frac{1}{n-w+1})$.
    
    Technically, we have constructed a \textbf{coupling} between the two families $(x_i)_{i∈[m-w+1]}$ and $(x_i')_{i∈[m-w+1]}$ of random variables, i.e. we have embedded them in the same probability space. The outcomes of $(x_i)_{i ∈ [m-w+1]}$ and $(x_i')_{i ∈ [m-w+1]}$ are now tightly linked. If $H ≥ α$ then we have $x_i ≥ x_i'$ for all $i$ and if $H ≤ α$ then we have $x_i ≤ x_i'$ for all $i$, and both happen with probability roughly $1/2$. By changing $α$ very slightly we can ensure that one of the two cases occurs whp. Relying on some monotonicity property of our surrounding problem such as “additional keys can only reduce the probability that all keys can be placed” allows us to \textbf{transfer a result that holds for the Poisson model to the Binomial setting}.
\end{digression}

With this change, the values $q₀,q₁,q₂,…$ from (\ref{eq:m/d/1-queue}) are the states of a so-called \textbf{M/D/1} queue, which is a \textbf{Markov chain} that can be illustrated like this:
\begin{quote}
    \begin{minipage}{\textwidth}
        \includegraphics[width=0.9\textwidth]{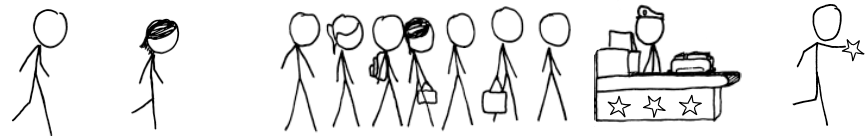}%
        \smash{\raisebox{-3pt}{\xkcdCredit{779}}}
        \nopagebreak[4]\\[-\baselineskip]
        \begin{minipage}[t]{3.5cm}\centering\mycaptionstyle
            $\underbrace{\hspace{\linewidth}}$\\
            \oldtextbf{M}arkovian arrivals,
            i.e.\ people arrive randomly with a \emph{rate} of $α$
        \end{minipage}\quad
        \begin{minipage}[t]{4.2cm}\centering\mycaptionstyle
            $\underbrace{\hspace{\linewidth}}$\\
            state of the process:\\number $q$ of people waiting
        \end{minipage}
        \begin{minipage}[t]{2.8cm}\centering\mycaptionstyle
            $\underbrace{\hspace{\linewidth}}$\\
            \oldtextbf{D}eter\-mi\-nis\-tic server handling \oldtextbf{1} person per time unit
        \end{minipage}
        \begin{minipage}[t]{1.5cm}\centering\mycaptionstyle
            $\underbrace{\hspace{\linewidth}}$\\
            served people leave
        \end{minipage}
    \end{minipage}
\end{quote}
If $α < 1$, then the customers arrive, on average, slower than they can be served and there is a \textbf{stable distribution} for the number $q^*$ of waiting customers in the long run.
\textbf{Queuing theory} literature promises
an average queue length of $𝔼[q^*] = 𝒪(1/(1-α))$ \cite{Cooper:QueingTheory-3rd:1990} and the tail bound $\Pr[q^* ≥ w] = \e^{-Θ(w/(1-α))}$ \cite[Prop 3.4]{EZB:M/D/1-tails:2006}. This means we have to pick $w = Ω(\log(n)/(1-α))$ to ensure that the size of $Q$ never exceeds $w$ within $𝒪(n)$ steps whp, which means all keys are validly placed in bounded linear probing.

\paragraph{So is this any good?}
We have essentially reinvented a \textbf{linear probing} hash table with the twist that the \textbf{maximum probe length} is guaranteed to not exceed $w = 𝒪(\log(n)/(1-α))$ when the load factor is $\frac{n}{m} ≈ \frac{n}{m-w+1} = α$. This is strongly related to \emph{Robin Hood hashing} \cite{Celis:RobinHood:1986} and not too exciting at first.

The arguments just given can be strengthened to show that the matrix from (\ref{eq:1-block-matrix}) with block length $w$ has independent rows whp.\footnote{The corresponding variant of the M/D/1 queue involves customers that randomly pay attention only half the time when they are in the queue. At every step, the left-most customer that pays attention is served, if any. When the queue is long, the speed at which customers are served is hardly affected. The additional time a customer spends in the queue is $𝒪(\log(n))$ whp.}
Moreover, a corresponding linear system can be solved with $𝒪(n/(1-α))$ row operations in expectation using \textbf{Gaussian elimination} because after sorting the rows by starting position the matrix is \textbf{already close to being in echelon form} (as already seen above). The number of row operations per column is linked to the average length of $Q$.
Assuming we can handle $𝒪(\log n)$ bits at a time using bit parallelism, the scalar product to be carried out by a query takes $𝒪(1/(1-α))$ operations. A \textbf{query} is very cache efficient because it reads $w$ \textbf{contiguous bits from memory}.

Overall we get a retrieval data structure with a load factor close to $1$ that needs to access one contiguous sequence of bits from memory per query, with no bullet biting concerning expensive linear system solvers. My coauthor Martin and I were quite pleased with this solution \cite{DW:One-Block-per-Row:2019}, which has since been dubbed \emph{ribbon retrieval} \cite{DHSW:Ribbon-SEA:2022}, because it improved upon the state of the art in 2019. Then we got an email from a database engineer who made it even better. \cite{dillinger:ribbon:2020}

\subsection{Bumped Ribbon: The Power of \texorpdfstring{$\bm{1+ε}$}{1+epsilon} Choices}

The following ideas were intended for retrieval, but here I continue in the simpler hash table setting, where they also apply.

\paragraph{More than one, less than two.} Let's take a step back. Consider a cuckoo hashing setting where each key is associated with $k$ random starting positions $s₁,…,s_k ∈ [m-w+1]$ and may be placed in cells with an index in $\bigcup_{i ∈ [k]} \{s_i,…,s_i+w-1\}$, i.e.\ within one of \textbf{$k$ randomly placed blocks of size $w$}.
\begin{center}
    \includegraphics[page=17]{img/table-pics.pdf}
\end{center}
For $k = 1$ we obtain the ribbon scheme discussed in \cref{sec:standard-ribbon} and for $k ≥ 2$ a scheme by Lehman and Panigrahy briefly mentioned in \cref{sec:hashing-scheme-gallery}. Let us now ask: What is the \textbf{smallest $w = w(k)$} such that all keys can be placed whp when the load factor is, say, $α = 0.99$?
We find:
\begin{center}
    \begin{tabular}{cccccccccc}
        \toprule
        $k$ & $1$ & $2$ & $3$ & $4$ & $5$ & $6$ & $7$ & …\\
        \midrule
        $w(k)$ & $Θ(\log n)$ & $3$ & $2$ & $2$ & $1$ & $1$ & $1$ &…\\
        \bottomrule
    \end{tabular}
\end{center}
The point is: There is a \textbf{qualitative difference} between having just one choice of a block (if you even want to call it “choice”) and the power of two or more choices. Could there be an interesting \textbf{middle ground between $1$ and $2$}?
\begin{center}
    \begin{minipage}{0.45\textwidth}
        \includegraphics[width=0.8\textwidth]{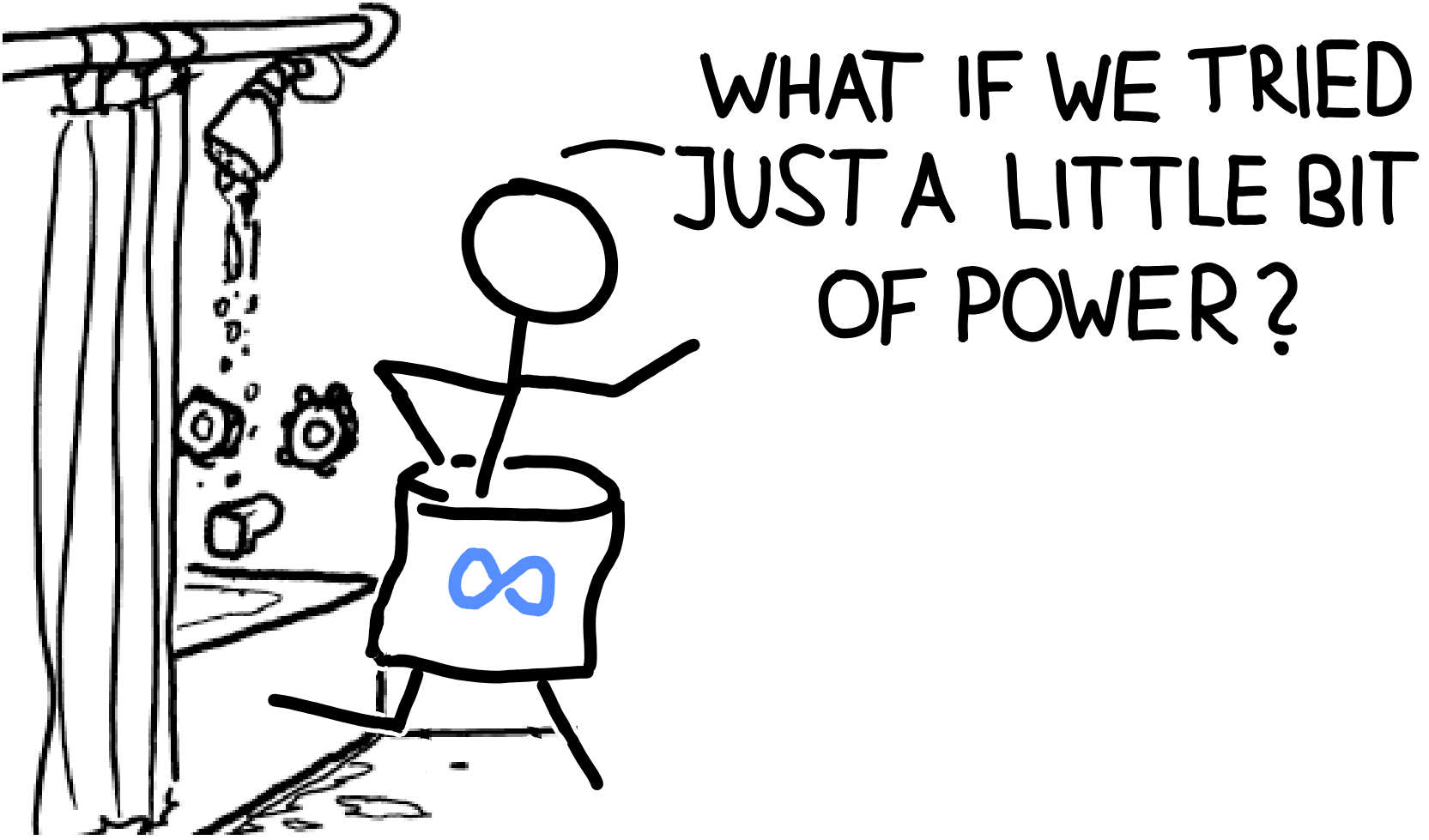}~
        \xkcdCredit{1715}
        \mycaptionstyle{How I imagine Peter Dillinger came up with a great idea.}
    \end{minipage}
\end{center}
Could there be a power of $k = 1.5$ choices? What would this even mean? Well, a key could have $2$ choices with probability $50\%$ and only $1$ choice with probability $50\%$. One might argue that this is $1.5$ choices per key on average.
I would nitpick and say that choices tend to aggregate multiplicatively (selecting among $2$ choices and then among $3$ choices gives $6$ choices overall) so taking the geometric mean and speaking of $\sqrt{2} ≈ 1.41$ choices per key is more natural.%
\footnote{This is the right view when considering the \emph{average} amount of \emph{information} that has to be recorded per key to encode the choices. This information in bits is the $\log₂$ of the number of choices.}
Either way, the problem is that placing the $≈ 50\%$ of keys that end up with $1$ choice (and ignoring the others) still requires $w = Ω(\log n)$ with no improvement over $k = 1$.

But here is a different kind of compromise. We partition the key set into groups of expected size $b$ and offer \textbf{$c$ choices for handling a group of keys} as a whole.
This (arguably) corresponds to $k = c^{1/b}$ independent choices per key.
Such a $k$ might well be strictly between $1$ and $2$. Dillinger et al.\ \cite{DHSW:Ribbon-SEA:2022} dubbed a concrete setup of this kind \emph{bumped ribbon}.

\paragraph{Bumped Ribbon.}
Like before, each key is associated with one block of $w$ consecutive positions. Keys are partitioned into groups with consecutive starting positions. For each group, there are two choices. The keys are either stored normally or the entire group is \textbf{\emph{bumped}}, meaning the keys are moved to a recursively constructed \textbf{fallback data structure}.\footnote{To bring this more in line with the previous setting, we can consider the fallback data structure to be a special segment in the primary data structure rather than separate and ensure that that bumped keys are not bumped \emph{again} by the fallback data structure. In \cite{DHSW:Ribbon-TR:2022} such a variant is called Bu${}¹$RR, but introducing it here would be a distraction from our main point.}
\begin{center}
    \includegraphics[width=0.5\textwidth]{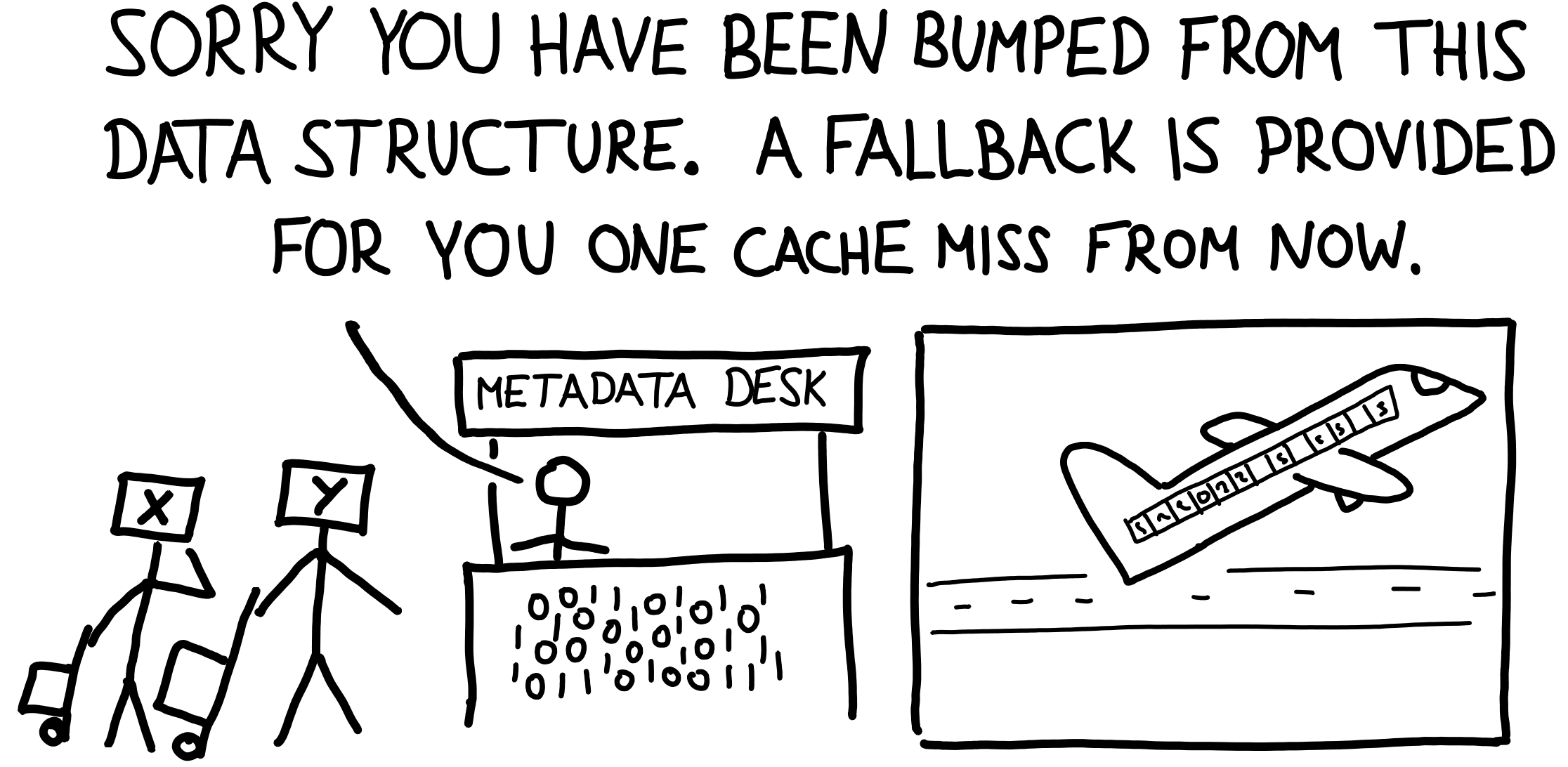}
\end{center}
To get some intuition, consider the following matrix-like visualisations were rows correspond to the keys sorted by starting position and columns to array cells. Grey shading in position $(i,j)$ indicates that the $i$th key may go into array slot $j$ (imagine we have zoomed out so the border of the grey area looks like a continuous curve).
We'd like for the \textbf{matrix diagonal} to run through the shaded region because then the $i$th key can go into the $i$th slot so
\textsf{(i)} every key would be placed
and \textsf{(ii)}
every slot would be put to use.
We can ensure \textsf{(i)} and \emph{mostly} ensure \textsf{(ii)} with two measures. We start with a \textbf{slightly overloaded} data structure (a load factor $α > 1$) and then \textbf{bump groups of keys in strategic positions}.
\begin{center}
    \begin{tabular}{ccc}
        \includegraphics[page=2]{img/ribbon-diagonal.pdf}
        &
        \raisebox{2cm}{$\longrightarrow$}
        &
        \raisebox{0.5cm}{\includegraphics[page=3]{img/ribbon-diagonal.pdf}}\\
        \begin{minipage}[t]{0.38\textwidth}\centering
            \mycaptionstyle
            an overloaded data structure where not all keys could be placed
        \end{minipage}
        &&
        \begin{minipage}[t]{0.38\textwidth}\centering
            \mycaptionstyle
            after bumping some keys, the remaining keys perfectly fill all cells
        \end{minipage}
    \end{tabular}
\end{center}

\begin{digression}[!b]{Some details on bumped ribbon.}
    \label{dig:bumped-ribbon}
    We are actually bumping keys by starting position, meaning a key is bumped if and only if its starting position is marked as bumped.
    The starting positions are partitioned into groups of size $𝒪(w²/\log w)$. For each group, we store 2 bits of metadata that indicate which positions are bumped: \textsf{(i)} none, \textsf{(ii)} the first $\frac{3}{8}w$ positions, or \textsf{(iii)} all positions, where \textsf{(iii)} is a rarely used emergency option. The initial load factor (before bumping) is set to be $α = 1+Θ(\frac{\log w}{w})$.
    The metadata can be set such that:
    \begin{enumerate}
        • Each non-bumped key is placed within $w$ cells of its starting position.
        • Only a $w^{-Ω(1)}$ fraction of the cells remains empty.
        • The overall memory overhead is dominated by the metadata and thus $𝒪(\frac{\log w}{w²})$ bits per key.
    \end{enumerate}
\end{digression}
Details on the exact proposal are given in \cref{dig:bumped-ribbon}.
For any $ε > 0$, bumped ribbon has $1+ε$ choices per key in the mean (for each \emph{group} of $𝒪(1/ε)$ keys, there are $3$ choices: bump none, some, or all keys from the group). The block size is $w = Ω({\log(1/ε)}/{\sqrt{ε}})$. The memory overhead (the fraction of wasted space taking into account unused cells, metadata and fallback data structures) is of order $𝒪(ε)$. In this sense, we are using a \textbf{power of $1+ε$ choices}!
Framed in terms of the previous table:
\begin{center}
    \begin{tabular}{ccccccccccc}
        \toprule
        $k$ & $1$ & $\bm{1+ε}$ & $2$ & $3$ & $4$ & $5$ & $6$ & $7$ & …\\
        \midrule
        $w(k)$ & $Θ(\log n)$ & $\bm{𝒪\big(\frac{\log(1/ε)}{\sqrt{ε}}\big)}$ & $3$ & $2$ & $2$ & $1$ & $1$ & $1$ & …\\
        \bottomrule
    \end{tabular}
\end{center} 

While we have not evaluated the performance of bumped ribbon as a static \emph{dictionary}, the corresponding static \emph{retrieval} data structure has significantly improved upon the state of the art by marrying three useful properties: First, courtesy of $1+ε$ choices, the \textbf{block size $w$ need not grow with $n$}. Second, there is a good chance of answering a query with a \textbf{single memory access} since most keys are not bumped. Third, we get \textbf{close to linear time construction} since the relevant linear system is already close to echelon form.

\DeclareRobustCommand{\tldr}{\begin{tikzpicture}[baseline=(v.base)]
    \node[inner sep=0.3] (v) {TL;DR};
    \draw[very thick] (v.south west) -- (v.north east);
\end{tikzpicture}}

\addcontentsline{toc}{section}{Conclusion}
\section*{\tldr\ Conclusion}


Our journey began with a discussion of cuckoo hash tables that have worst-case access times of $𝒪(1)$ and a load factor of $α = \frac 12 - ε$.
We considered various approaches for increasing the load factor and argued that increasing the number of independent choices that every key has is not the most attractive option due to the increased number of cache misses incurred by each query. While two choices per key is qualitatively different from a single choice (i.e.\ “no choice”), the step to more than two choices is comparatively underwhelming. What's more, when considering static retrieval data structures closely related to cuckoo hash tables we find that the flexibility afforded by multiple choices can come at the price of expensive construction algorithms.

Wondering if “two” is really the least amount of choice one can have per key motivates \emph{bumped ribbon}. When used as a hash table it is much like linear probing, except that the maximum probe length is guaranteed to never exceed a constant $w$. Keys can be marked as “bumped”, meaning they are placed in a fallback data structure, if their part of the hash table is too crowded. Superficially, there are two choices per key: bumped or not bumped. However, there is more rigidity in two important ways: First, only few keys are bumped so the entropy in any key's choice is much less than one bit. Second, bumping decisions are correlated with large groups of keys being bumped as a whole.

In a sense we have harnessed the power of $1+ε$ choices per key. The qualitative power of choices is present, namely the ability to defuse certain worst-case constellations, but the price that flexibility brings is significantly attenuated.


\bibliographystyle{plainurl}
\bibliography{bibliographie}

\end{document}